\documentclass[prl,aps,twocolumn,showpacs,floatfix,superscriptaddress,showkeys]{revtex4}
\usepackage{epsfig}
\usepackage{bm}
\usepackage{times}
\usepackage{mathptm}
\usepackage{mathptmx}
\def\days{\text{83}}
\def\users{\text{3,188}}
\def\usersm{\text{3,187}}
\def\links{\text{309,125}}
\def\nonmass{\text{202,695}}
\def\pairs{\text{7,087}}
\def\directedlinks{\text{20,879}}
\def\pairmess{\text{105,349}}
\def\cutoff{\text{0.5}}
\def\curvature{\text{0.1}}
\def\minemail{\text{10}}

\def\ab{\sim}
\begin{document}
\title{Dialog in e-Mail Traffic}
\author{Jean-Pierre Eckmann}
\affiliation{D\'epartement de Physique Th\'eorique,
Universit\'e de Gen\`eve, 1211 Gen\`eve 4,
Switzerland}
\affiliation{Section de Math\'ematiques,
Universit\'e de Gen\`eve, 1211 Gen\`eve 4,
Switzerland}
\author{Elisha Moses}
\affiliation{Department of Physics of Complex Systems,
Weizmann Institute of Science, Rehovot 76100, Israel}
\author{Danilo Sergi}
\affiliation{D\'epartement de Physique Th\'eorique,
Universit\'e de Gen\`eve, 1211 Gen\`eve 4,
Switzerland}
\date{\today}
\begin{abstract}
Connectivity and topology are known to yield information about networks, whose origin is
self-organized,  but the impact of {\it temporal dynamics} in a
network is still mostly unexplored. Using an information theoretic approach to e-mail exchange, we show that an
e-mail network allows for a separation of static and dynamic structures within it. The static
structures are related to organizational units such as departments. The temporally linked
structures turn out to be more goal-oriented, functional units such as committees and user
groups.
\end{abstract}
\keywords{e-mail network; synchronization; thematic organization}
\maketitle
The theory of complex networks has developed tools to investigate quantitatively the
properties of a number of new systems, such as the World Wide Web (WWW), the protein-protein 
interaction database and others (see
\cite{JPMoses,Barabasirep} and references therein for a review).  The challenging difficulty
shared by such systems is the interplay between their constituents, and the resulting
collective effects.  When the system is a fixed graph whose links describe interaction,
concepts such as the clustering coefficient (or curvature) have been introduced and applied
successfully. In e-mail networks {\em temporal dynamics} appears as a new ingredient, and
therefore it is possible to ask about the synchronization of e-mail traffic between
communicating  users, and to determine the correlation (cohesion) between them.

In this Letter, we show how to obtain an  objective measure of the interaction between the
activity of users by employing tools provided by information theory, or the general theory of
entropy
\cite{CT}. We find that the temporal structure of the e-mail exchange reveals a new form of
organization that is different from what can be captured by the more static notion of
curvature, or any other study which neglects temporal aspects. It is intriguing that the
question of how an organization communicates internally is similar to those that arise in the
study of how an organ like the brain organizes its neuronal activity. We find analogies in
our analysis to the approach which studies the appearance of correlations and synchronization
between spike trains in the creation of a neural code
\cite{spike}.

\paragraph{The experiment} Our data are extracted from the log files of one of the main mail
servers at one of our universities, and consist of over $2\cdot10^6$ e-mail messages sent
during a period of $\days$ days, connecting about ten thousand users.  The content of the
messages is of course never accessible, and the only data taken from the log file are the
`to', `from', and `time' fields. The data are first reduced to the internal mail within the
institution, since external links are necessarily incomplete. Once aliases are resolved, we
are left with a set of $\users$ users interchanging
$\links$ messages.

A directed graph is then constructed by designating users as nodes and connecting any two
of them with a directed link if an e-mail message has gone between them during the $\days$
days. This procedure defines a {\it static} graph. Statistical properties of the
degree of this graph have been reported before
\cite{Kiel,ShowerLaws}. Connectivity of this {\it static} graph will reveal structures
within the organization
\cite{Kiel,self,Hubermann}. We have previously shown
\cite{JPMoses} that a powerful tool for identifying such structural organization is the
number $t$ of triangles (triplets in which all pairs communicate) that a node
of valence
$v$ (total number of partners) participates in, normalized by the number of triangles
${v(v-1)/ 2}$ that it could potentially belong to. This defines the clustering coefficient
$c={2t/v(v-1)}$, which as we showed induces a curvature on the graph \cite{JPMoses}.

One marked difference between the graphs of e-mail and of the WWW should be noted at this
point. In the WWW, the central organizing role of `hubs' (nodes with many outgoing links)
that confer importance to `authorities' (nodes with many ingoing links) has been noted
\cite{Kleinberg} and utilized very successfully (e.g., by Google). The contribution of
authorities and hubs is, however, {\em not} to the creation of communities and interest
groups. This is evident since the high valence of both hubs and authorities tends to reduce
their curvature considerably. High curvature nodes, in contrast, are usually the specialists
of their community, that are highly connected in bi-directional links to others in the group.
In the e-mail graph, hubs tend to be machines, mass mailers or users that transfer general
messages (e.g., seminar notifications),  going out to many users, while authorities are more
like service desks. Thus the importance of hubs and authorities is small if we consider the
core use of the e-mail structure as dealing with thematic rather than organizational issues.
They do, however, play a role in such questions as diffusion of viruses, or more generally,
how many people are being reached
\cite{Kiel}. But most mass mailings do not solicit an answer, and therefore do not contribute
to interaction (`dialog') as we define and study it in this Letter. In our analysis we
discard mass mailings (more than $18$ recipients) altogether. There remain
$\nonmass$ links.

The different manner in which triangles and transitivity interplay in the WWW and in the
e-mail graphs is also illuminating. The notion of curvature is a local one, based on the more
basic concept of a `co-link'. This is a link between two nodes that point to each other,
establishing a `friendly' connection based on mutual recognition. Building from the single
pair, the fundamental unit of connectivity is the triangle \cite{Uri}. In the WWW
transitivity is natural, and we have shown previously that if node $A$ is `friendly' with
nodes
$B$ and $C$, it is often correct to assume that $B$ and $C$ are friends. On the contrary,
e-mailing  is so prolific that $A$'s having a dialog with $B$ and with $C$ usually does not
imply that $B$  and $C$ carry out a dialog, and even if they do then the three communications
determining the edges of the triangle could be independent; as a consequence, transitivity
breaks down. We will see that static structures (such as departments) emerge as
high-curvature ones, while dialog between members of a group implies a more functional, and
perhaps goal-oriented structure.

Our analysis expands the notion of a mutual, or `co-link' to the e-mail network by
designating a link between nodes $A$ and $B$ only if $A$ has sent a message to
$B$ and {\it $B$ has sent a message back} during the whole period under investigation. We find
$\pairs$ such pairs, sending $\pairmess$ messages to each other, among
$\directedlinks$ directed pairs who sent perhaps mail just one way (and out of the
${\users\cdot
\usersm/ 2}$ possible connections in the graph).


\paragraph{The model}
To analyze the behavior of this reduced network, we view any pair of `conversing'  users as
exchanging signals on a transmission line on which information can be propagated in both
directions. We completely disregard the fact that there is internal information in the
messages, discarding even information that is in principle available in the log files such as
the size of the messages.  The data for each pair is a spike train whose horizontal axis is
time, with upward ticks for a mail sent $A\to B$ and a downward tick for $B\to A$ (some
samples are shown in Fig.~\ref{im:fig2}). We now define that {\bf
$A$ and $B$ conduct a dialog on a given day} if $A$ sends mail on that day to
$B$ and
$B$ answers {\em on the same day}.

\begin{figure}[hbtp]
\begin{center}
\includegraphics[width=0.4\textwidth]{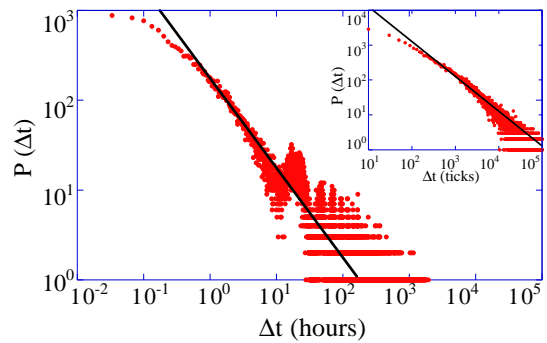}
\caption{The probability distribution of the response time till a message is `answered' (see
text for definitions). Inset: same but
measured in `ticks', i.e. units of messages sent in the system. Solid lines follow
$\sim \Delta t^{-1}$ and are meant as a guide to the eye. }
\label{im:fig1}
\end{center}
\end{figure}

\begin{figure*}[hbtp]
\begin{center}
\includegraphics[width=\textwidth]{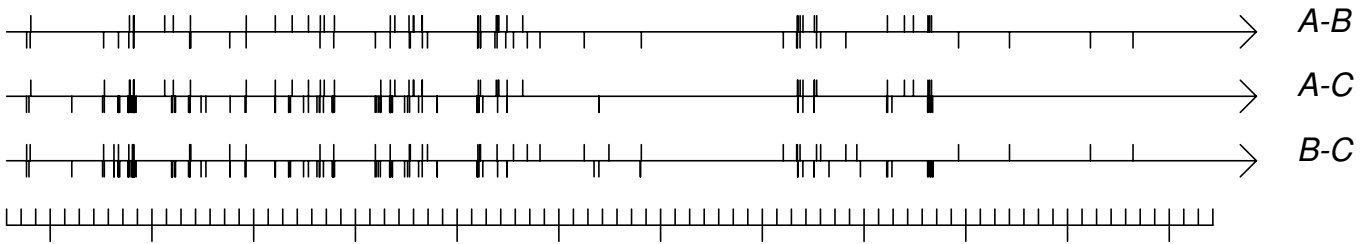}
\caption{Spike trains for the three communication channels determining the
  edges of a triangle formed by the users $A$, $B$ and $C$. We have
  $I_{p}(A,B)=0.095$, $I_{p}(A,C)=0.394$, $I_{p}(B,C)=0.172$ and
  $I_{t}=1.606$. It is important to note that $I_{p}$ and $I_{t}$
  capture the synchronization of the e-mail exchange at two different
  levels. $I_{t}$ measures the coherence of the triangle as a whole, and
  can take on high values even though some of the $I_{p}$'s are
  relatively small. The fourth horizontal line represents the whole
  period under analysis divided into days (and weekends), introduced with the purpose
  to help to visualize the events determining the probabilities
  entering $I_{p}$ and $I_{t}$ (cf.~text).}
\label{im:fig2}
\end{center}
\end{figure*}

The temporal dynamics of the e-mail network immediately reveals new statistical properties,
shown in Fig.~\ref{im:fig1}. We define $\Delta T$ as the time delay between a message going
from
$A\to B$ and a response going from $B\to A$. While no clear power law is evident in
Fig.~\ref{im:fig1}, the behavior can be approximated by $P(\Delta t) \approx
\Delta t^{-1}$.  The appearance of a peak ranging from
$\Delta t = 16$ to $\Delta t = 24$ can be explained by sociological behavior involving the
time (usually $16$ hours) between when people leave work and when they come back to their
offices. This (already very weak) peak disappears when considering in the inset the basic
time unit as a `tick' of the system ($=$ a message sent). We suspect that the approximate
power law is caused by random communications between two users, while the flat incipient part
implies actual correlation between two users (when the answer comes before
$10$ hours have passed, i.e. on the same day).

Choosing the basic `tick' of the clock (the sending of a message in the network) as a
{\it variable} time unit smoothens many features (as in Fig.~\ref{im:fig1}, inset). In
particular, the slowing down of the network over nights and weekends is eliminated. But the
mathematics of `correlation' becomes much more involved, and we have also checked that the
interaction is very well captured by sticking to the more intuitive notion of `same day'.
In light of these considerations, we choose 24 hours as the natural time unit within which
$B$ is required to answer. In principle, some multiple of this unit could serve as well,
but the results of Fig.~\ref{im:fig1} show that most interactions take place within 10
hours. We further checked that extending the choice of a day to an answer on the next day
gives similar results.

The mathematical description of the experiment proceeds through two steps,
cf.~Fig.~\ref{im:fig2}. First, at a more local level, we consider a pair of communicating
users, that we shall denote by
$A$ and $B$. We introduce the probabilities $p_{A}(i)$ and $p_{B}(i)$, where
$i=0,1$. The value
$1$ corresponds to the event that at least one e-mail has been sent to the partner on a given
day, while the value $0$ corresponds to having sent none on that day. The measured values of
these probabilities are given by
\[p_{A}(i)=N_{A}(i)/ {d}~,\] where $N_A(i)$ is the number of days for which the event $i$
occurred for $A$, and similarly for $B$ (and $d$ is the total number of days
$d=\days$). We then characterize the {\em joint} activity of $A$ and $B$ by considering the
probabilities $p_{AB}(i,j)$ defined as
\[p_{AB}(i,j)=N_{AB}(i,j)/d~,\] where $N_{AB}$ is the number of days where $A$ was in state
$i$ and
$B$ in state $j$ (i.e., sending mail to the partner or not) and
$i,j\in\{0,1\}$. It is now possible to determine to which extent the activity of $A$
influences the activity of $B$ by means of the {\em mutual information}
$I_{p}(A,B)$ (the subscript $p$ stands for pair):
\[I_{p}(A,B)=\sum_{i,j=0,1}p_{AB}(i,j)\cdot\log\bigg(\frac{p_{AB}(i,j)}{p_{A}(i)\cdot
p_{B}(j)}\bigg).\]
$I_{p}$ measures in what way knowing what $A$ does will predict what
$B$ does and vice versa (note that $I_{p}(A,B)=I_{p}(B,A)$).

The next step consists in considering every triangle of communicating users; to be specific
we designate them by $A$, $B$, and $C$. In order to capture their joint activity we introduce
the probabilities
$p_{ABC}(i_{1},i_{2};i_{3},i_{4};i_{5},i_{6}) = p_{ABC}({\bf i})$, where
$i_{1},\dots , i_{6}\in\{0,1\}$. The pair $(i_{1},i_{2})$ refers to the communication
$A\leftrightarrow B$,
$(i_{3},i_{4})$ to $A\leftrightarrow C$, and $(i_{5},i_{6})$ to
$B\leftrightarrow C$. For example the pair
$(i_{1}=1,i_{2}=0)$ has to be interpreted as the occurrence where on a given day $A$ sends
mail to $B$, but $B$ does not send mail to $A$. An equivalent, evident interpretation holds
for all other pairs. In formulas the above probabilities read
\[p_{ABC}(i_{1},i_{2};i_{3},i_{4};i_{5},i_{6})=N_{ABC}(i_{1},i_{2};i_{3},i_{4};i_{5},i_{6})/d~,\]
$N_{ABC}({\bf i})$ being the number of days where the pattern (event) ${\bf i}$ occurred.

\begin{figure}[hbtp]
\begin{center}
\includegraphics[width=0.4\textwidth]{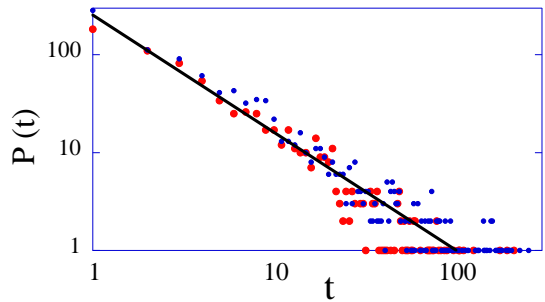}
\caption{Histogram of static and temporal statistical quantities. Probability distribution of
the number
$t$ of triangles that a user participates in. Blue circles indicate static
triangles, while red ones indicate `temporally cohesive' triangles (i.e. mutual
information $I_t \ge 0.1$). Both lines are well fit by $\ab \Delta t^{-1.2}$ and the black line
is a guide to the eye with this slope.}
\label{im:fig1new}
\end{center}
\end{figure}

We now define the {\em temporal cohesion} of a triangle as the degree of
synchronization between the activity of the three users. This is achieved by looking at a
form of the {\em mutual information} $I_{t}(A,B,C)$ (in this case the subscript
$t$ stands for triangle) defined as
\begin{eqnarray} I_{t}(A,B,C)&=&\sum_{i_{1},\dots, i_{6}=0,1} p_{ABC}({\bf i})\nonumber\\
&\times&\log\bigg(\frac{p_{ABC}(i_{1},i_{2};i_{3},i_{4};i_{5},i_{6})}{p_{AB}(i_{1},i_{2})\cdot
  p_{AC}(i_{3},i_{4})\cdot p_{BC}(i_{5},i_{6})}\bigg).\nonumber\\\nonumber
\end{eqnarray} Note that the temporal cohesion
$I_{t}(A,B,C)$ is invariant under any permutation of $A$, $B$ and $C$. Also,
$I_t\le\log(16)$ and the maximum is attained when the four possible patterns for each edge
are equiprobable and fully correlated. More insight into $I_{p}$ and
$I_{t}$ can be gained by looking at Fig.~\ref{im:fig2}, showing the three communications
determining a triangle.  A statistical quantity of interest, shown in Fig.~\ref{im:fig1new}
is the number $t$ of triangles that a user participates in. The distribution of both
the static and the dynamic (temporal cohesion $\ge 0.1$) triangles follow a power law over
two decades, with exponent $-1.2$.

\paragraph{Restoring transitivity} With the help of the temporal cohesion
$I_{t}$ it is now possible to replace the static transitivity by a novel notion of temporal
transitivity. The assumption is that if the  e-mail exchange in a triangle is highly
synchronized, the three users are indeed involved in a common dialog. This transitive
relationship between users can be extended naturally to adjacent triangles. This idea relies
on the observation that in the presence of two highly synchronized triangles with a common
edge, the four users are supposed to influence each other's activity. In this way it is
possible to extract the groups of users carrying out a dialog. We thus construct a new,
conjugate, graph where we first draw a node for each triangle for which $I_{p}$ is larger 
than a given cutoff.  Two of these nodes will be connected by a link if the corresponding
triangles  have a common edge, that is, if 4 people $A$, $B$, $C$, $D$ are involved in these
2 triangles (say $A$, $B$, $C$, and $B$,
$C$, $D$).  Such a construction (called the conjugate graph) will offer a perspective on the
appearance of circles of users sharing a common interest, defining thematic groups.

\begin{figure}[hbtp]
\begin{center}
\includegraphics[width=0.4\textwidth]{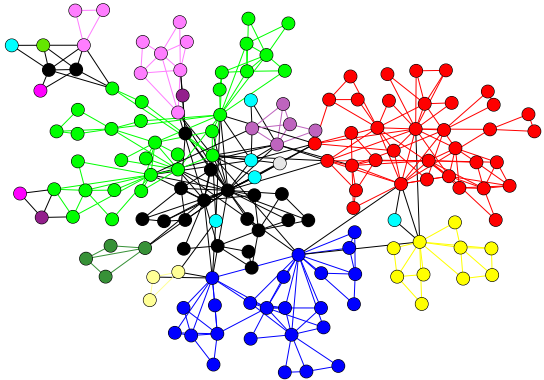}
\caption{The static structure of the graph of e-mail traffic, arranged according to
curvature, based on triangles of mutual recognition. Time is thus {\em not} taken into
account, and the graph of users arranges itself primarily according to departments, shown in
various colors.}
\label{im:fig3}
\end{center}
\end{figure}

\paragraph{Discussion of the results} For the purpose of comparison, we first consider in
Fig.~\ref{im:fig3}  the {\em static} graph resulting from our e-mail network. For the sake of
clarity only nodes, i.e. users, with a curvature larger than
$\curvature$ are  present; in addition, every pair of users must have exchanged at least
$\minemail$  e-mails. The temporal dynamics intrinsic to the e-mail exchange is here
neglected and triangles represent a sign of static transitive recognition, carrying no
information about temporal cohesion between the individual communications. In this case we
see the clear appearance of departmental communities. Our findings on the organizational
aspects of the e-mail traffic are thus in agreement with the findings of
\cite{Hubermann}, but are based here on the quantitative concept of curvature.

\begin{figure}[hbtp]
\begin{center}
\includegraphics[width=0.4\textwidth]{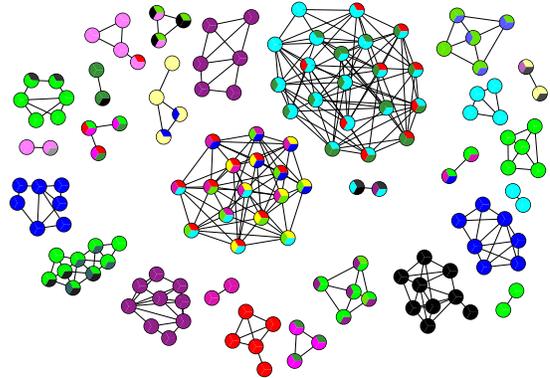}
\caption{The conjugate graph, for a cutoff of $\cutoff$. Each node is
  a triangle of 3 people conferring with temporal cohesion $I_t\ge\cutoff$, and
each link
  connects two adjacent such triangles. The 3 colors of each node are
  the departments of the 3 people (same color code as in
  Fig.~\ref{im:fig3}).
  Note the strong clustering of
  the graph into very compact groups of people. The users {\em cross}
  department boundaries (their interests and connections are not shown
  out of considerations of privacy).}
\label{im:fig4}
\end{center}
\end{figure}

Fig.~\ref{im:fig4}, on the other hand,  shows the conjugate graph associated with a cutoff of
$I_t\ge\cutoff$. We recognize several highly connected, totally separated clusters,
indicating different thematic groups. Some of the clusters identified in Fig.~\ref{im:fig2} survive and are lifted in part to the graph of temporal dynamics, indicating
indicating that within some departments there are dialogs; furthermore 
some departments split into 
different interest groups.
However, we find
many clusters that are new, and do not appear in the high curvature graph. These are
typically users that are not in the same department, as shown by the multiple colors of the
disks. Very few users appear in more than one cluster, so that the spreading of functional
information is restricted within the thematic communities, in contrast to spreading of
computer viruses for example, which propagate easily through the entire graph
\cite{Kiel}.

Some conclusions may be drawn regarding the nature of communities that emerge
by conducting a dialog in the internet network.  Two people engaged in a
project can, if necessary, pick up the phone and tie all loose ends
efficiently. However, a group with three or more participants may find it hard
to coordinate conference calls, and in general will benefit from the lower time
constraints that allow each participant to formulate his views and present
them to a forum by e-mail. This makes e-mail an ideal medium for discussion
groups involved in a given project, or a committee involved in a functional
activity. Indeed, we have identified two committees in the clusters of
Fig.~\ref{im:fig4} that are involved in non-academic activities within the
university. A third group can be identified as visiting scientists (e.g.
post-docs etc.) from a common foreign nationality.

The choice of a university's e-mail network is perhaps not ideal for
identifying such `groups of dialog'. This is because the major activity in a
university is research, which usually involves few individuals, and is almost
never advanced by committee. We thus speculate that the role of dialog in
defining functional communities will be greater in large organizations such as
companies \cite{Hubermann} or government offices.
\paragraph{Summary} We have studied e-mail communications over $\days$ days and quantified
the synchronization of groups of coherently communicating users. This synchronization reveals
the existence of common interests within those groups. This form of organization cannot
emerge by applying the analysis based on static concepts such as curvature, since those
detect only {\em structural} rather than {\em thematic} organization. The reason is that in
the context of static e-mail networks a triangle does not automatically imply transitivity
from a thematic point of view. But we have demonstrated that transitivity can be re-captured
by taking into account the temporal dynamics of the e-mail traffic.

\begin{acknowledgments} We thank P. Collet for many helpful
discussions and  P. Choukroun and A. Malaspinas for help with the
e-mail data.
This work was
supported by the Fonds National Suisse, and by the Minerva Foundation (Munich).
\end{acknowledgments}


\begin{thebibliography}{2}
\bibitem{ShowerLaws} We have confirmed a scaling law for the number $M_S$ of messages
sent and received ($M_R$) per user, and find that both scale as $M_{S,R}\sim M_{S,R}^{-1.0
\pm 0.1}$. However, the degree of the graph, i.e. the valence or number of links going in
$v_{\text{in}}$ and out
$v_{\text{out}}$ of a node, does not have a simple scaling behavior. Instead, one power law
of $v^{-0.93}$ characterizes the behavior of both $v_{\text{in}}$ and
$v_{\text{out}}$ till
$v=12$, where a sharp transition to a second power law with
$v^{-1.84}$ fits the behavior till $v\approx 200$ where our data end.
\end{thebibliography}
\end{document}